\documentclass[aps,prl,reprint,groupedaddress,amsmath,amssymb]{revtex4-1}
\usepackage[final]{graphicx}
\usepackage{verbatim}
\usepackage{color}

\begin{document}

\title{First Results from a Microwave Cavity Axion Search at $24~\mu\text{eV}$}

\author{B.~M.~Brubaker}\email{benjamin.brubaker@yale.edu}
\author{L.~Zhong}
\author{Y.~V.~Gurevich}
\author{S.~B.~Cahn}
\author{S.~K.~Lamoreaux}
\affiliation{Department of Physics, Yale University, New Haven, Connecticut 06511, USA}
\author{M.~Simanovskaia}
\author{J.~R.~Root}
\author{S.~M.~Lewis}
\author{S.~\surname{Al Kenany}}
\author{K.~M.~Backes}
\author{I.~Urdinaran}
\author{N.~M.~Rapidis}
\author{T.~M.~Shokair}
\author{K.~A.~\surname{van Bibber}}
\affiliation{Department of Nuclear Engineering, University of California Berkeley, Berkeley, California 94720, USA}
\author{D.~A.~Palken}
\author{M.~Malnou}
\author{W.~F.~Kindel}
\author{M.~A.~Anil}
\author{K.~W.~Lehnert}
\affiliation{JILA and the Department of Physics, University of Colorado and National Institute of Standards and Technology, Boulder, Colorado 80309, USA}
\author{G.~Carosi}
\affiliation{Physics Division, Lawrence Livermore National Laboratory, Livermore, California 94551, USA}

\date{\today}

\begin{abstract}
We report on the first results from a new microwave cavity search for dark matter axions with masses above 20~$\mu\text{eV}$. We exclude axion models with two-photon coupling $g_{a\gamma\gamma} \gtrsim 2\times10^{-14}~\text{GeV}^{-1}$ over the range $23.55 < m_a < 24.0~\mu\text{eV}$. These results represent two important achievements. First, we have reached cosmologically relevant sensitivity an order of magnitude higher in mass than any existing limits. Second, by incorporating a dilution refrigerator and Josephson parametric amplifier, we have demonstrated total noise approaching the standard quantum limit for the first time in an axion search.
\end{abstract}

\maketitle

\textit{Introduction.}---Astrophysical and cosmological measurements over the past few decades overwhelmingly favor a $\Lambda$CDM cosmology in which more than 80\% of the matter in the universe is nonrelativistic, nonbaryonic ``dark matter'' whose particulate nature remains unknown~\cite{planck2016}. The axion is a hypothetical particle predicted by the Peccei-Quinn solution to the Strong $CP$ problem~\cite{PQ1977a,*PQ1977b,weinberg1978,*wilczek1978}, and sufficiently light axions are also excellent cold dark matter candidates, with extremely weak couplings to standard model fields~\cite{pww1983,*as1983,*df1983}. Historically, $1~\mu\text{eV} \lesssim m_a \lesssim 1~\text{meV}$ has been cited as the allowed mass range for dark matter axions, with $10 < m_a < 50~\mu\text{eV}$ preferred~\cite{rmp2003}. More recent lattice QCD calculations favor $m_a \gtrsim 50~\mu\text{eV}$~\cite{lattice2016}, subject to the usual uncertainties from early universe chronology.

Axions constituting the galactic halo may be detected in the lab via their Primakoff conversion into monochromatic microwave photons in a high-$Q$ cavity permeated by a strong magnetic field~\cite{sikivie1985,krauss1985}. The Axion Dark Matter eXperiment (ADMX) has refined this technique since 1996 and ruled out narrowband power excesses $\gtrsim 10^{-22}~\text{W}$ over 3~K noise between 460 and 892~MHz, thus excluding a range of viable axion models with $1.9 < m_a < 3.69~\mu\text{eV}$~\cite{ADMX1998,*ADMX2002,*ADMX2004,*ADMX2010,*ADMX2016,[{See }][{ for a recent review of axion detection more generally.}]graham2015}. To date these are the only dark matter axion limits with cosmological sensitivity; technologies facilitating detection at higher masses are thus urgently needed.

In this Letter, we report the first results from a new microwave cavity detector sited at the Yale Wright Laboratory. By pushing to lower temperatures and leveraging tremendous recent progress in quantum electronics, we have set the first limits with cosmologically relevant sensitivity above $20~\mu\text{eV}$ axion mass. These are also the first cavity results at any frequency to approach the fundamental noise limits imposed by quantum mechanics, and thus demonstrate a technical achievement crucial to the full exploration of axion parameter space.

\textit{Detection principle.}---A cavity axion detector consists of a tunable microwave cavity coupled to a low-noise receiver, maintained at cryogenic temperature in the bore of a high-field magnet. The conversion power is enhanced when $m_ac^2/h\simeq\nu_c$, where $\nu_c$ is the resonant frequency of a cavity mode with an appropriate spatial profile. Exactly on resonance, the signal power in natural units is
\begin{equation}
P_S = \left(g_\gamma^2\frac{\alpha^2}{\pi^2}\frac{\rho_a}{\Lambda^4} \right)\left(\omega_cB_0^2VC_{mn\ell}Q_L\frac{\beta}{1+\beta}\right).
\label{eq:power}
\end{equation}

In this expression the first set of parentheses contains the theory parameters: $\alpha$ is the fine-structure constant, $\rho_a\simeq0.45$~GeV/cm$^{3}$ is the local dark matter density~\cite{[{}][{. Both 0.45~GeV/cm$^{3}$ (also used by ADMX) and the more commonly cited 0.3~GeV/cm$^{3}$ fall within the range of recent measurements.}]read2014}, $\Lambda=78$~MeV encodes the dependence of the axion mass on hadronic physics, and $g_\gamma$ is a model-dependent dimensionless coupling. Two models denoted KSVZ~\cite{kim1979,*SVZ1980} and DFSZ~\cite{DFS1981,*zhitnitskii1980}, with $g_\gamma=-0.97$ and $0.36$ respectively, have historically served as useful benchmarks for experiments. But more accurately KSVZ and DFSZ are both families of models, for which $\left|g_\gamma\right|$ can be as large as 4.6 or as small as 0.03~\cite{cheng1995,kim1998}; experiments probing this ``model band'' are properly described as cosmologically sensitive. The physical coupling that appears in the axion-photon Lagrangian is $g_{a\gamma\gamma}=\left(g_\gamma\alpha/\pi\Lambda^2\right)m_a$.

\begin{figure*}[t]
\includegraphics{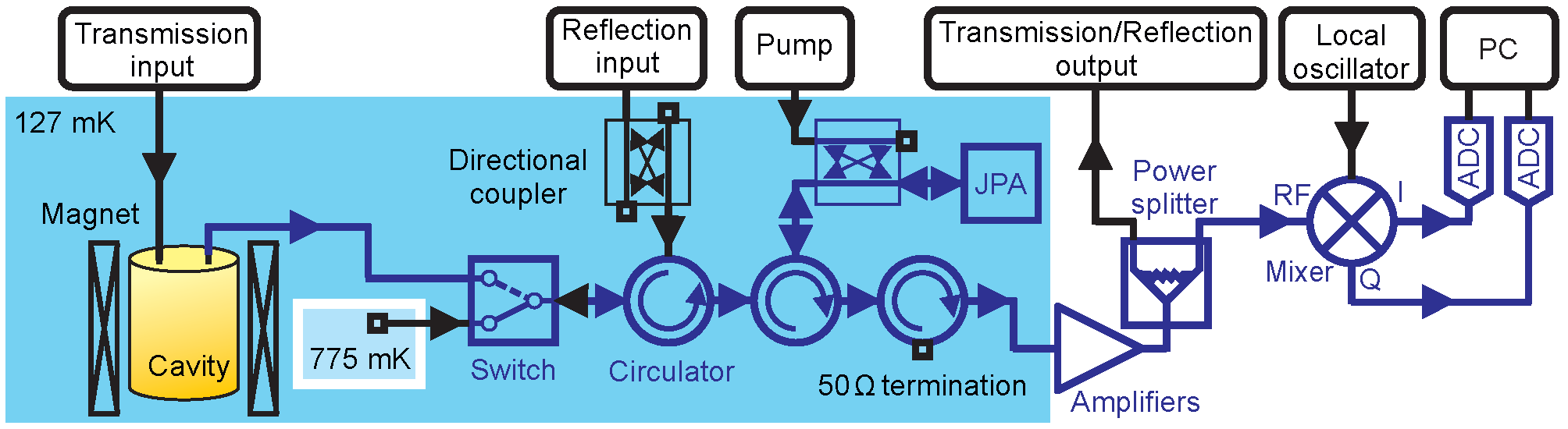}
\caption{\label{fig:diagram} Simplified receiver diagram. Blue arrows indicate the path that a putative axion signal would take through the system, and black arrows indicate other paths. A vector network analyzer (VNA) is used to measure the cavity's frequency response in both transmission and reflection.}
\end{figure*}

The factors in the second set of parentheses in Eq.~\eqref{eq:power} are properties of the detector, where $B_0$ is the magnetic field strength, $V$ is the cavity volume, and $\omega_c=2\pi\nu_c$. The mode's coupling to the receiver, parametrized by $\beta$, reduces the quality factor from $Q_0$ to $Q_L=Q_0/\left(1+\beta\right)$, and the form factor $C_{mn\ell}$ quantifies the overlap between the static $B$ field and the mode's $E$ field. For cylindrical cavities $C_{mn\ell}$ is only appreciable for the lowest-order TM$_{0n0}$ modes.

Inserting typical values for our detector, $P_S\simeq 5\times10^{-24}$~W on resonance for a KSVZ axion with $m_a=24~\mu\text{eV}$. The signal power inherits the Maxwellian functional form of the standard isothermal halo energy spectrum, with velocity dispersion $\left<v^2\right>^{1/2}\simeq 270~\text{km/s}$. The axion signal linewidth is thus $\Delta\nu_a = m_a\left<v^2\right>/h \simeq 5$~kHz for $m_a\simeq24~\mu\text{eV}$, much smaller than a typical cavity linewidth $\Delta\nu_c=\nu_c/Q_L \simeq 500$~kHz.

The axion mass is \textit{a priori} unknown, so we tune the cavity in discrete steps $\lesssim\Delta\nu_c/2$. If we average the cavity noise for a time $\tau$ at each step, the resulting signal-to-noise ratio (SNR) is
\begin{equation}
\Sigma = \frac{P_S}{k_BT_S}\sqrt{\frac{\tau}{\Delta\nu_a}}.
\label{eq:SNR}
\end{equation}
Assuming a phase-insensitive linear receiver, the system noise temperature $T_S$ is given by
\begin{equation}\label{eq:noise}
k_BT_S = h\nu\left(\frac{1}{e^{h\nu/k_BT} -1} + \frac{1}{2} + N_A\right),
\end{equation}
where the three additive contributions correspond, respectively, to a blackbody gas in equilibrium with the cavity at temperature $T$, the zero-point fluctuations of the blackbody gas, and noise added by the receiver. Phase-insensitive linear receivers are subject to quantum limits that enforce $N_A \geq 1/2$~\cite{caves1982}, from which we obtain the standard quantum limit $k_BT_S \geq h\nu$ for microwave cavity axion detection. At each step the detector is sensitive over a bandwidth of about $2\Delta\nu_c$, so several consecutive steps will contribute to the SNR at each frequency.

\textit{Experimental design.}---Our detector (discussed in greater detail in Ref.~\cite{NIM2016}) is housed in a cryogen-free dilution refrigerator integrated with a 9 T superconducting solenoid from Cryomagnetics, Inc. The cavity hangs in the magnet bore on a gantry anchored to the dilution refrigerator's mixing chamber plate, which is maintained at $T_C=127$~mK by electronic feedback.

Our cavity is a 2~L copper-plated stainless cylinder whose TM modes may be tuned by rotation of a copper rod occupying 25\% of the cavity volume. The rotation is mechanically driven by a stepper motor at room temperature. Two other stepper motors control the insertion of a coaxial antenna and a thin dielectric rod into the cavity, used for adjusting $\beta$ and fine-tuning, respectively. The results reported in this Letter were obtained using the TM$_{010}$ mode between 5.7 and 5.8~GHz. Typical parameter values in this range are $Q_0\simeq3\times10^4$ and $C_{010}\simeq0.5$; we set $\beta\simeq2$ to optimize the scan rate~\cite{NIM2016}.

Our receiver (Fig.~\ref{fig:diagram}) was designed to both minimize the system noise and enable robust and flexible \textit{in situ} noise calibration. We realize these goals by incorporating a near-quantum-limited Josephson parametric amplifier (JPA) and a microwave switch at the receiver input. The switch may be toggled between the cavity and a 50~$\Omega$ termination thermally anchored to the dilution refrigerator's still plate at $T_H=775$~mK; this arrangement allows us to interleave noise calibrations into the axion search.

Our preamplifier comprises the JPA itself (described below) as well as a directional coupler for the JPA's microwave pump input and a circulator to separate input and output signals. Two other circulators are required to isolate the JPA from both the cavity and the second-stage amplifier, a high electron mobility transistor (HEMT) at 4~K. At room temperature, the signal is amplified further, down-converted to an intermediate frequency (IF) band centered at 780~kHz, and digitized. 

The JPA is essentially a nonlinear $LC$ circuit that owes its inductance to an array of Superconducting Quantum Interference Devices (SQUIDs)~\cite{castellanos2007}. Parametric amplification occurs when the JPA is driven with a strong pump tone near its resonant frequency, which can be tuned by varying the DC flux through the SQUID array. The JPA gain profile is always centered on the pump frequency, with peak gain and bandwidth determined by the pump power and flux bias; at our $21$~dB operating gain, the bandwidth is $2.3$~MHz. A multilayer system comprising a bucking coil, passive persistent coils, and both ferromagnetic and superconducting shields is used to reduce the external field at the SQUID array to $\sim10^{-4}$~G. Even with this shielding, slow flux drifts on long time scales can compromise the JPA gain, so we stabilize the flux using a feedback system which maximizes the power in a weak tone injected near the JPA resonance. 

When the spectrum of the signal to be amplified is symmetric about the pump tone, the JPA amplifies the signal quadrature in phase with the pump and deamplifies the other. This configuration evades the quantum limit cited above, but does not improve the SNR without a second JPA and added operational complexity~\cite{zheng2016}. Here we operate the JPA with the cavity resonance and all Fourier components of interest detuned to the high-frequency side of the pump tone. In this configuration the JPA acts as a phase-insensitive amplifier, subject to $N_A \geq 1/2$.

When we calibrate the receiver's added noise at frequencies far detuned from the cavity mode, the cavity looks like an open circuit, and the cold load noise comes from a terminated port on the directional coupler in the reflection input line (see Fig.~\ref{fig:diagram}). In this configuration we obtain $N_A=1.35$~quanta in total, most likely due to 0.63 quanta from the vacuum and thermal contributions to the JPA's added noise at $T_C$, approximately 0.2 quanta of HEMT noise referred to the JPA input, and approximately 0.5 quanta from about 2~dB of loss in microwave components before the JPA. 

Noise calibrations near the cavity mode indicate a roughly Lorentzian excess in the cold load noise with a peak value around 1~quantum, which we attribute to a poor thermal link between the tuning rod and the cavity barrel. Thus the total noise (this excess cavity thermal noise plus the three terms in Eq.~\eqref{eq:noise}, with $N_A=1.35$) is $k_BT_S\simeq3h\nu$ on resonance in each spectrum, falling to $2.2h\nu$ at 650~kHz detuning in either direction.

\textit{Operations.}---We acquired axion search data from January 26 to March 5 and again from May 16 to August 2, 2016. The full data set consists of about 7000 measurements from two long scans across the full range and several shorter scans to compensate for nonuniform tuning.

Acquisition of this data was fully automated and controlled by a LabVIEW program. At each iteration, this program tunes the TM$_{010}$ mode, extracts $\nu_c$, $Q_L$, and $\beta$ from vector network analyzer (VNA) measurements of the cavity, then sets the local oscillator frequency to $\nu_c + 780$~kHz and the JPA pump frequency to $\nu_c - 820$~kHz. It then adjusts the JPA pump power and flux bias to optimize the gain, turns on the flux feedback system, and samples both IF channels to collect $\tau=15$~min of axion-sensitive data. Power spectra are constructed and averaged in parallel with time-stream data acquisition, with image rejection implemented in software in the frequency domain, resulting in a single heavily averaged spectrum with bin width $\Delta\nu_b=100$~Hz. The noise calibration is repeated every ten iterations; the overall live-time efficiency is 72\%. We thus make \textit{in situ} measurements of every parameter appearing in Eqs.~\eqref{eq:power} and \eqref{eq:SNR} that can change between iterations, with the exception of $C_{010}$, whose frequency dependence is obtained from simulation.

During our first full scan, we injected synthetic axion signals with $\Delta\nu \simeq 5$~kHz through the transmission line at ten random frequencies, with a nominal intracavity power of $10^{-22}$~W. A factor of two uncertainty on this power level, due to unknown cryogenic insertion losses of individual components, prevents us from independently calibrating the sensitivity using fake signal injections. Nonetheless, such injections are still valuable as a fail-safe check on the data acquisition and analysis procedures; we found power excesses $>5\sigma$ at all the expected frequencies in the combined first scan data.

\begin{figure}[t]
\includegraphics{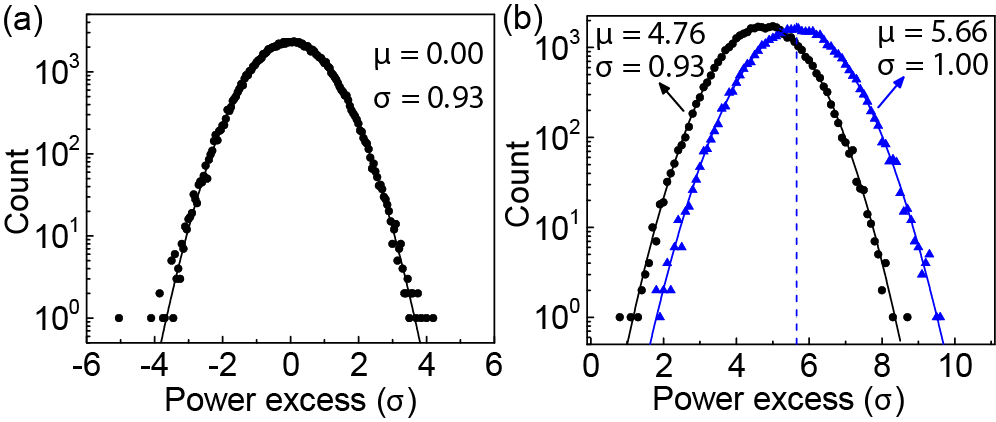}
\caption{\label{fig:hist} (a) Histogram of the full grand spectrum constructed from real data after cutting synthetic axions. (b) Simulated axion signal in the presence of Gaussian white noise. We construct the simulated grand spectrum two ways: multiplying the noise by an empirical baseline and applying a Savitzky-Golay filter to mimic real data ($\bullet$), and using an ideal flat baseline and no fitting ({\color{blue} $\blacktriangle$}). In both cases the single grand spectrum bin best aligned with the axion is histogrammed over many iterations of the simulation. The reduction in $\sigma$ is due to the fit-induced correlations also observed in real data. After correcting for the reduction in $\sigma$, the reduction in $\mu$ is the fit-induced axion power loss. The vertical line is an analytic calculation of the expected SNR.}
\end{figure}

\begin{figure*}[t]
\includegraphics{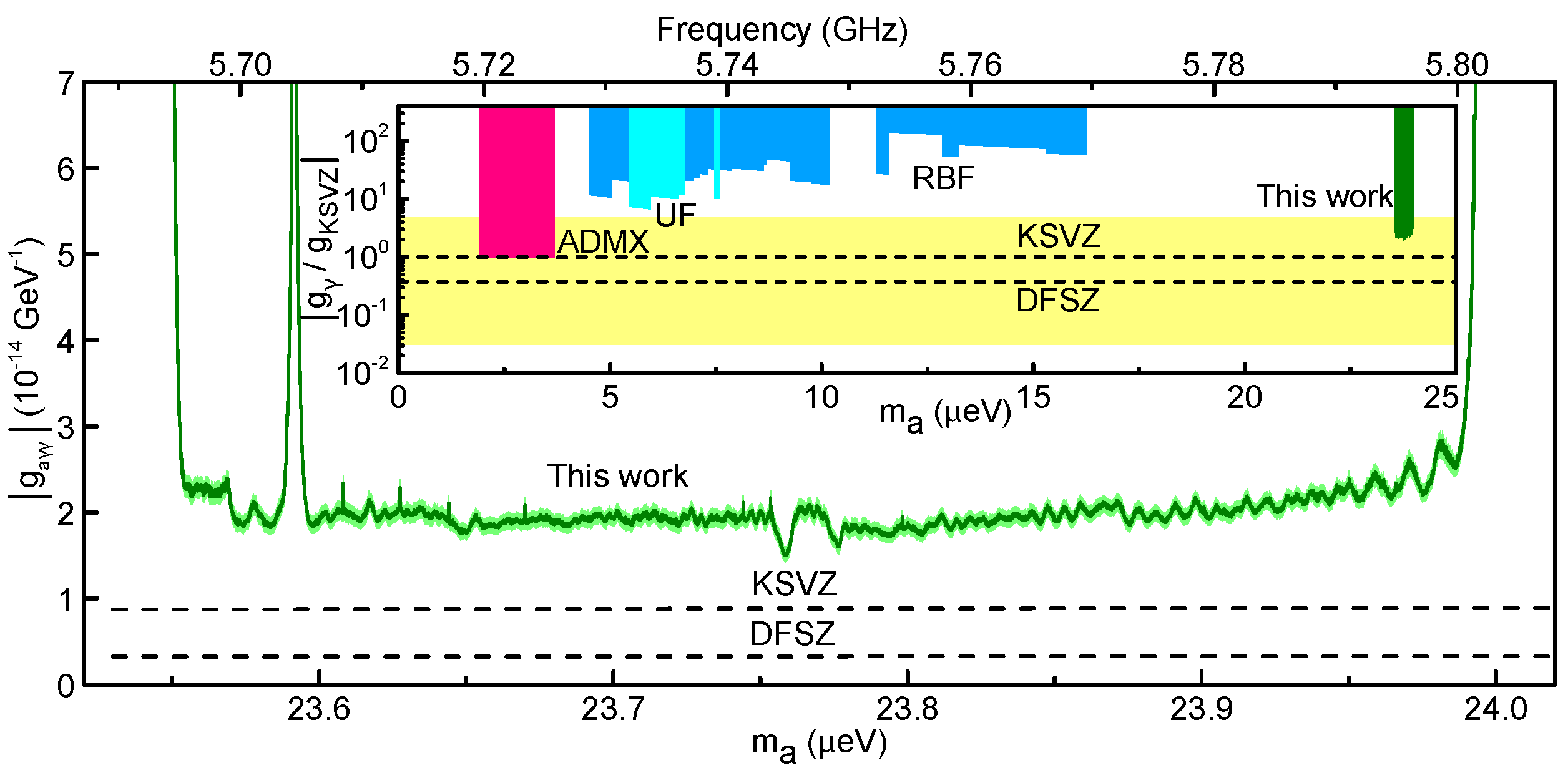}
\caption{\label{fig:plot} Our exclusion limit at 90\% confidence. The light green shaded region is a $1\sigma$ error band. The large notch around 5.704 GHz is the result of cutting spectra around a previously unidentified TE mode. The narrow notches correspond to frequencies where synthetic axion signals were injected in one of the scans. The inset shows this work (green) together with previous cavity limits from ADMX (magenta, Ref.~\cite{ADMX2010}) and early experiments at Brookhaven (RBF, blue, Ref.~\cite{RBF1987,*RBF1989}) and the University of Florida (UF, cyan, Ref.~\cite{[{}][{. For uniformity of presentation, both RBF and UF limits have been rescaled to $\rho_a = 0.45$~GeV/cm$^3$ from their original published values where  $\rho_a = 0.30$~GeV/cm$^3$ was used. Note also that both RBF and UF used 95\% confidence limits.}]UF1990}). The axion model band~\cite{cheng1995} is shown in yellow.}
\end{figure*}

\textit{Analysis.}---We restrict all analysis to an IF band of full width $1.3~\text{MHz} \gtrsim 2\Delta\nu_c$ centered on the cavity mode at 780~kHz. Our procedure follows the basic model proposed in Ref.~\cite{ADMX2001} with various refinements to be discussed more thoroughly in a forthcoming publication. We first average all the spectra to extract the average shape of the spectral baseline in the analysis band; this averaging also reveals individual channels contaminated by single-bin IF features, which are cut from the subsequent analysis. After we divide out the average baseline, individual spectra still exhibit 1~dB residual variation on large spectral scales. We fit the residual variation in each spectrum with a Savitzky-Golay (SG) filter, which is most usefully regarded as a digital low-pass filter with a very flat passband~\cite{schafer2011}. By dividing out the SG fit and subtracting 1 we obtain a set of Gaussian white noise spectra representing excess power which we call the ``processed spectra.'' In the absence of axion conversion each bin in each processed spectrum is a sample drawn from the same Gaussian distribution, with $\mu=0$ and $\sigma = 1/\sqrt{\Delta\nu_b\,\tau} = 3.3\times10^{-3}$.

In the presence of axion conversion, we expect the mean power to be nonzero (but still $\ll \sigma$) in about 50 consecutive bins in each of the processed spectra in which the frequency corresponding to the axion mass appears. We construct a combined spectrum whose value in each RF bin is given by a sum of the corresponding IF bins across all processed spectra, weighted according to their different sensitivities. More precisely, the weights are chosen to yield the maximum likelihood estimate for the mean power in each combined spectrum bin, and we normalize each bin to the maximum-likelihood weighted quadrature sum of sample standard deviations from the contributing processed spectra. The probability density function of the combined spectrum at 100 Hz resolution is Gaussian with $\mu=0$ and $\sigma=1$, as we expect. 

We then sum nonoverlapping 10-bin segments of the combined spectrum to reduce the resolution to 1~kHz, and construct a ``grand spectrum'' whose $i$th bin is a weighted sum of the $i$th through $(i+4)$th 1~kHz bins. The weights are chosen to yield the maximum likelihood mean power in each grand spectrum bin assuming a Maxwellian axion lineshape with velocity dispersion $\left<v^2\right>^{1/2} = 270~\text{km/s}$; each bin is normalized to its expected standard deviation as above. Thus, we expect a Gaussian probability distribution for the grand spectrum with $\mu=0$ and $\sigma=1$. The actual distribution is histogrammed in Fig.~\ref{fig:hist}(a): it is Gaussian with the correct mean but $\sigma=0.93$. We have demonstrated via simulation that the reduction of $\sigma$ is due to the finite stop-band attenuation of the SG filter, which leads to small negative correlations between nearby 100 Hz bins. Because we understand the origin of these correlations, we can correct for their effects on the statistics of the grand spectrum.

In each grand spectrum bin the SNR at any constant coupling $\left|g_\gamma\right|$ can be computed as a quadrature sum of terms with the form of Eq.~\eqref{eq:SNR}, weighted according to the axion lineshape. We must also insert signal attenuation factors that do not appear in Eq.~\eqref{eq:SNR}. The SG fit will attenuate any real axion signal for the same reason that it reduces $\sigma$ on 5~kHz scales; the results of a simulation to quantify this fit-induced power loss are plotted in Fig.~\ref{fig:hist}(b). We also consider loss due to misalignment of the axion signal relative to the grand spectrum binning and loss before the microwave switch to which the noise calibration is not sensitive: the product of all three loss factors is $\eta=0.76$. We then adjust the coupling in each bin to obtain a constant target SNR; the resulting frequency-dependent coupling $\left|g_\gamma(\nu)\right|$ is the final value used to set an exclusion limit.

We chose a $5.1\sigma$ SNR target, corresponding to a candidate threshold of 3.455$\sigma$ at 95\% confidence. There were 28 grand spectrum bins exceeding this threshold, consistent with the candidate yield from simulated Gaussian white noise subjected to the same processing. We rescanned these 28 candidates in August 2016, with sufficient exposure to obtain high SNR at the appropriate coupling for each candidate frequency. Again we set a threshold at 95\% confidence, and none of the original 28 frequencies recurred as a candidate. We can therefore claim $\left|g_\gamma(\nu)\right|$ as an exclusion limit at $\geq90\%$ confidence. On average we exclude $\left|g_\gamma\right| \gtrsim 2.3\times \left|g_\gamma^{\,\text{KSVZ}}\right|$ for $23.55 < m_a < 24.0~\mu\text{eV}$; the corresponding limit on the physical coupling, $\left|g_{a\gamma\gamma}\right| \gtrsim 2\times10^{-14}~\text{GeV}^{-1}$, is plotted in Fig.~\ref{fig:plot}, with a 4\% uncertainty dominated by the calibration of the excess cavity thermal noise~\footnote{Uncertainty in $C_{010}$ is not well quantified and thus was not included in this error budget, but good agreement between simulation and preliminary field profiling measurements suggests an error $\lesssim10\%$}.  

\textit{Conclusion.}---Thirty years after microwave cavity detection of dark matter axions was first proposed, it remains the only technique with proven sensitivity to cosmologically relevant couplings. Until now cavity experiments have only achieved this sensitivity at the lowest allowed axion masses, primarily because the effective volume $VC_{mn\ell}$ falls off rapidly with increasing frequency. In this work we have demonstrated that despite this unfavorable scaling, a sufficiently low-noise experiment can reach the model band above $20~\mu\text{eV}$. We reported on the first successful operation of an axion detector incorporating a dilution refrigerator and JPA. These innovations resulted in total noise within a factor of 3 of the standard quantum limit, and an exclusion limit $\left|g_\gamma\right| \gtrsim 2.3\times \left|g_\gamma^{\,\text{KSVZ}}\right|$ over a 100~MHz frequency range. Further operation of this detector in the next few years will extend this frequency coverage significantly, and ongoing cavity and amplifier development~\cite{lamoreaux2013,NIM2016} may even enable us to push down to KSVZ sensitivity. 

\begin{acknowledgments}
This work was supported by the National Science Foundation, under grants PHY-1362305 and PHY-1306729, by the Heising-Simons Foundation under grants 2014-181, 2014-182, and 2014-183, and by the U.S. Department of Energy through Lawrence Livermore National Laboratory under Contract DE-AC52-07NA27344. We gratefully acknowledge the critical contributions by Matthias B\"{u}hler of Low Temperature Solutions UG to the design of and upgrades to the cryogenic system.
\end{acknowledgments}


%

\end{document}